\documentclass[pra,twocolumn,showpacs,floatfix,amsmath,amsfonts]{revtex4}

\newcommand{\cU}{{\cal U}}
\newcommand{\cG}{{\cal G}}
\newcommand{\cE}{{\cal E}}

\usepackage{epsfig}
\usepackage{graphicx}
\usepackage{amsmath}
\usepackage{amssymb}
\begin{document}
\title{Delocalizing transition in one-dimensional condensates in optical lattices  due to inhomogeneous interactions}
\author{Yu. V. Bludov$^1$}
\email{bludov@cii.fc.ul.pt}
\author{V. A. Brazhnyi$^1$}
\email{brazhnyi@cii.fc.ul.pt}
\author{V. V. Konotop$^{1,2,3}$}
\email{konotop@cii.fc.ul.pt}

\affiliation{
$^1$Centro de F\'{\i}sica Te\'orica e Computacional,
Universidade de Lisboa, Complexo Interdisciplinar, Avenida Professor Gama
Pinto 2, Lisboa 1649-003, Portugal
\\
$^2$Departamento de F\'{\i}sica,
Universidade de Lisboa, Campo Grande, Ed. C8, Piso 6, Lisboa
1749-016, Portugal \\
$^3$Departamento de Matem\'aticas, E. T. S. de Ingenieros Industriales, Universidad de Castilla-La Mancha 13071 Ciudad Real, Spain }

\pacs{03.75.Lm, 03.75Kk, 03.75Ss}

\begin{abstract}

It is shown that inhomogeneous nonlinear interactions in a Bose-Einstein condensate loaded in an optical lattice can result in delocalizing transition in one dimension, what sharply contrasts to the known behavior of discrete and periodic systems  with homogeneous nonlinearity. The transition can be originated either by decreasing the amplitude of the linear periodic potential or by the change of the mean value of the periodic nonlinearity. The dynamics of the delocalizing transition is studied.

\end{abstract}

\maketitle

\section{Introduction}

Spatial localization of the energy is a fundamental physical phenomenon which attracts a great deal of attention in very different physical
applications and mathematical statements. When localized states are originated by the balance between nonlinearity and natural dispersion of the
medium, they are referred to as solitons. If localization is caused by the interplay between nonlinearity and periodicity, induced either by
variations of parameters  of the medium or by discreteness, the respective solutions are usually termed localized modes (or gap solitons). Localized
modes are known to exist in numerous systems possessing the two main ingredients -- nonlinearity and periodicity. Discrete systems~\cite{disc-sys},
photonic crystals~\cite{Optics}, Bose-Einstein condensates (BECs) loaded in optical lattices (see e.g. ~\cite{BK,OM}) are only few examples where the
localized modes have been discovered.

In this context BECs in optical lattices have attracted a special attention, just after they have been created experimentally~\cite{Kasevich}. This
is due to flexibility of optical lattices allowing one to change their parameters in time and in space by simple management of geometry and intensity
of laser beams, as well as due to possibilities of experimental realization of one--, two--, and three--dimensional (below 1D, 2D, and 3D,
respectively) periodic systems. Naturally, a question about survivals of localized states subjected to variation  of the lattice parameters, or in
other words about existence of delocalizing transition has raised. This issue has been addressed within the framework of the discrete nonlinear
Schr\"{o}dinger equation~\cite{KRB}, which models the Gross-Pitaevskii (GP) equation in terms of the Wannier function expansion~\cite{AKKS}, and
directly within the framework of the GP equation with a  periodic potential~\cite{BS}. The most important finding of the carried research can be
formulated as the existence of the delocalizing transition in 2D and 3D cases and absence of such transition in 1D systems. The phenomenon was
explained~\cite{BS} by the fact, that in 1D lattices there exists no threshold for exciting localized states, which at small amplitudes become matter
gap-solitons (see e.g.~\cite{KS1,BK,OM}). Such a threshold appears in 2D and 3D systems. As it was mentioned in~\cite{KRB}, this explanation well
corroborates with analogous phenomenon known for discrete systems, where existence of discrete envelope solitons of arbitrarily small amplitudes is
possible in the 1D case~\cite{env-sol} but in two dimensions requires a threshold intensity~\cite{FKM}.

In this paper we show that, in a sharp contrast to the described situation, {\em in the case of  periodic nonlinearity delocalizing transition is
possible in a 1D system}. This transition, resulting from the interplay between dimensionality and nonlinear periodicity,  is studied in detail within the framework of the 1D GP equation and can be induced by change of either the lattice
potential or the nonlinearity. Focusing on the application of the results to the mean-field theory of condensed atomic gases, we notice that
periodically varying nonlinearity appears in description of BEC in optical lattices which are subjected to optically or magnetically
induced Feshbach resonance. Then a nonlinear lattice can be created by applying an optical standing wave periodically modulating the scattering length~\cite{Shliap}, by illuminating the regions of the linear lattice minima by focused laser beams introducing local changes of the scattering length, or by a spatially periodic magnetic field which can be produced by an array of superconducting or ferromagnetic materials~\cite{BEC-per}.
Periodical nonlinearity appears also in a natural way when one considers a boson-fermion mixture in an optical lattice provided a number of fermions
significantly exceeds the number of bosons. In that case the stationary Thomas-Fermi distribution of the  spin polarized fermions modifies the linear lattice and creates an effective nonlinear lattice for bosons due to the interspecies interactions, as this was discussed in detail in Refs.~\cite{BLK,SBKK}.

The organization of the paper is as follows. In Sec.~\ref{sec:model} we introduce the model and describe the physical arguments supporting existence of the delocalizating transition.  In Sec.~\ref{sec:transition} we discuss how the parameters of the system should be chosen and describe details of the dynamics of the delocalization. Sec.~\ref{sec:nl_trans} is devoted to a new mechanism of the delocalizing transition, which is based on the change of the nonlinearity controlling parameter (which is referred simply as a number of particles). The outcomes are summarized in Conclusion.

\section{The model}
\label{sec:model}

To explain the physics of the phenomenon we start with  the dimensionless 1D GP equation
\begin{eqnarray}
\label{psi_fin} i\psi_t=-\psi_{xx}+\cU(x)\psi+\cG(x)|\psi|^2\psi,
\label{boson1D}
\end{eqnarray}
where $\cU(x)$ is a $\pi$-periodic potential, $\cU(x)=\cU(x+\pi)$, and  $\cG(x)$ is the coordinate-dependent
nonlinearity, which is also $\pi$-periodic: $\cG(x)=\cG(x+\pi)$. In such a scaling the energy is measured in the units of the recoil energy $E_r=\hbar^2\pi^2/(2m d^2)$, where $d$ is the lattice constant, and the spatial and temporal variables in the units $d/\pi$ and $\hbar/E_r$, respectively. Particular physical units of $\cG(x)$ depend on the application of the model (see e.g.~\cite{BEC-per,BLK}).

Eq. (\ref{psi_fin}) possesses stationary localized mode solutions~\cite{SM,BLK,Chik}, $\psi=\phi(x)e^{-i\mu t}$. Here $\mu$ is a chemical potential. It must belong either to a finite $n$-th gap ($n=1,2...$): i.e. $\mu\in  \left({\cal E} _n^{(-)},\cE_{n}^{(+)}\right)$, where $\cE_n^{(\sigma)}$ is either lower ($\sigma$ stands for "$-$")
or upper ($\sigma$ stands for "$+$") edge of the gap, or to the semi-infinite gap $\mu\in \left(-\infty, \cE_0^{(+)}\right)$ of the spectrum of the operator ${\cal L}=-d^2/dx^2+\cU(x)$, i.e.
 ${\cal L}\varphi_n^{(\sigma)}=\cE_n^{(\sigma)}\varphi_n^{(\sigma)}$, where $\varphi_n^{(\sigma)}(x)$ is a normalized Bloch state: $\int_{0}^{\pi}|\varphi_n^{(\sigma)}(x)|^2dx=1$.
All branches of such solutions are parameterized by the chemical potential. They either collapse at some values of $\mu$ inside a gap or end up at a
gap edge. The lowest branch may  bifurcate from the extended Bloch state. If this happens, the mode represents a small amplitude gap soliton governed
by the nonlinear Schr\"odinger equation (see e.g.~\cite{KS1,BK})
$
    iA_\tau=-\left(2M_n^{(\sigma)}\right)^{-1}A_{\xi\xi}+\chi_n^{(\sigma)}|A|^2A
$
for a slow envelope amplitude $A(\tau,\xi)$, which is introduced through the representation $\psi\approx\epsilon A(\tau,\xi)\varphi_n^{(\sigma)}(x)$
and depends on slow variables  $\xi=\epsilon x$ and $\tau=\epsilon^2 t$. The small parameter $\epsilon$ is proportional to  the energy detuning
towards the gap from the edge $\cE_n^{(\sigma)}$; $\epsilon\sim|\mu-\cE_n^{(\sigma)}|$, $M_n^{(\sigma)} =\left[d^2\cE_n^{(\sigma)}/dk^2\right]^{-1}$
is the effective mass and $\chi_n^{(\sigma)}=\int_{0}^{\pi}\cG(x)|\varphi_n^{(\sigma)}(x)|^4dx$ is the effective nonlinearity.

In a general situation, when the lattice amplitude decreases, the width of a gap decreases as well and all gaps collapse at $\cU(x)\equiv 0$. In its
turn a small gap may allow for the existence of only small amplitude solitons which tend to extended Bloch waves in the limit when the chemical
potential approaches the gap edge. Existence of the solitons prevents delocalizing transition, because adiabatic decrease of the lattice amplitude
leads to increase of the soliton width and subsequent increase of the potential results in recovering a localized state~\cite{KRB,BS}.

Thus, in order to obtain delocalizing transition one has to create a situation where small amplitude excitations do not exist. This is, in particular, the case of 2D and 3D lattices~\cite{KRB,BS,FKM}. In order to explore such a possibility in the 1D model (\ref{psi_fin}) we recall that the condition for the modulational~\cite{KS1,BK} (also referred to as dynamical~\cite{dyn-instab}) instability now reads $M_n^{(\sigma)}\chi_n^{(\sigma)}<0$. For the
existence of small amplitude solitons this condition must be satisfied near at least one of the gap edges. This happens, in particular, in models with the homogeneous nonlinearity ($\cG\equiv$const)~\cite{KS1,BK} and in discrete models with on-cite nonlinearity~\cite{env-sol}. Thus, the requirements
\begin{eqnarray}
\label{crit}
M_n^{(+)}\chi_n^{(+)}>0\quad\mbox{and}\quad M_n^{(-)}\chi_n^{(-)}>0,
\end{eqnarray}
when satisfied {\it simultaneously}, represent the conditions for possibility of delocalizing transition. It is to be mention here that inequalities
(\ref{crit}) represent the constrains of nonexistence of stationary gap solitons, while other type of moving coupled-mode solitons can be excited in
shallow lattices when the gap width becomes small enough (such solitons are well known in nonlinear optics~\cite{Aceves} and for different condensed
atomic gases were recently reported in Ref.~\cite{SBKK,Fatkhul}).

The main difference of the problem at hand from the cases of GP equation either with a periodic potential and a homogeneous
nonlinearity~\cite{KS1,BK} or with periodic nonlinearity and zero potential~\cite{SM}, is that the obtained criterion (\ref{crit}) strongly depends
on the Bloch states and a specific form of the nonlinearity $\cG(x)$.

\section{Delocalizing transition}
\label{sec:transition}

\begin{figure}
  \begin{center}
   \begin{tabular}{c}
       \scalebox{1.0}[1.0]{\includegraphics{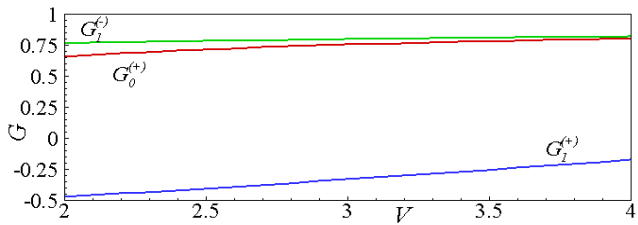}}
   \end{tabular}
   \end{center}
\caption{The lines  $G_{n}^{(\sigma)}(V)$ separate domains where $\chi_n^{(\sigma)}(V,G)>0$ and $\chi_n^{(\sigma)}(V,G)<0$ (above and below the respective lines).
} \label{fig:chi}
\end{figure}

 The signs of the effective mass at the different
gap edges are defined: $\mp M_n^{(\pm)}<0$. Hence, in order to satisfy (\ref{crit}) we have to check the signs of the effective nonlinearity
$\chi_n^{(\pm)}$. We consider a stable localized state in a semi-infinite gap, i.e. with a solution of (\ref{boson1D}) with $\mu<\cE_0^{(+)}$. To be
specific we explore the model with $\cU(x)=-V\cos(2x)$ and $\cG(x)=G-\cos(2x)$. The lattice amplitude $V$ and the "average" nonlinearity $G$ are used
to control signs of the effective nonlinearities $\chi_n^{(\sigma)}$. Let us denote by $G_{n}^{(\sigma)}$ the values of $G$ at which
$\chi_n^{(\sigma)}$ becomes zero, $\chi_n^{(\sigma)}=0$.  Since the effective nonlinearity is a functional of the Bloch states, it depends on lattice
amplitude $V$, i.e. $G_{n}^{(\sigma)}=G_{n}^{(\sigma)}(V)$. Then the curves $G_{n}^{(\sigma)}(V)$ on the plane $(V,G)$ separate domains where the
respective nonlinearities $\chi_n^{(\sigma)}$ acquire positive and negative values (for the semi-infinite and the first bands they are shown in
Fig.~\ref{fig:chi}).
 Consequently, in the semi-infinite gap, we have to look for the delocalizing transition  in the $(V,G)$-domain above $G_{0}^{(+)}$. In the first gap the transition can be expected for the parameters corresponding to the region above $G_{1}^{(+)}$ and below $G_{1}^{(-)}$.

\begin{figure}[h]
\epsfig{file=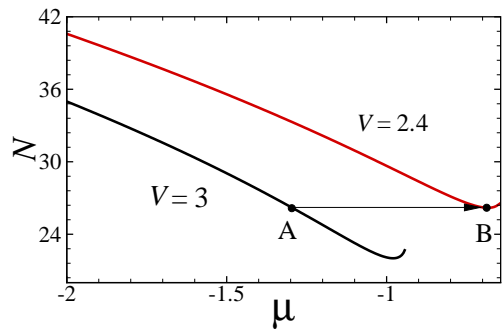,width=7cm} \caption{Number of particles $N$ {\it vs} chemical potential $\mu$ for the semi-infinite gap for $V=3$ and $V\approx
2.4$. The arrow AB corresponds to $N_A=26.25$.  } \label{fig:dyn_si_a}
\end{figure}

Let us now suppose that $G$ is fixed. Then,  to estimate the lattice amplitude $V_{thr}$ at which the transition occurs we first analyze the dependence of
the number of particles $N=\int_{-\infty}^{\infty}{|\psi(x)|^2dx}$ on the chemical potential. For the lowest
branch of the localized modes in the semi-infinite gap such dependencies are depicted in Fig.~\ref{fig:dyn_si_a} for different $V$. Each of the curves has a minimum whose location we denote as $(\mu_*,N_*)$. When the lattice potential decreases, the position of the branches on the plane $(\mu,N)$ is changed in such a way that its minimum, $N_*\equiv N_*(V)$, grows. Physically this means that for localization in more shallow lattices higher nonlinearities are necessary.

Let us assume that at some initial value $V_{ini}$ the chosen localized mode corresponds to the point A (see Fig.~\ref{fig:dyn_si_a}), i.e. it has a
number of atoms, which we denote $N_A$. Then, one can compute the threshold value $V_{thr}$, as one resulting in the minimum of the numbers of
particles to occur precisely at the same number as one of the original mode has, i.e. at $N_*(V_{thr})=N_A$ (in the figure this is indicated by the
point B). Designating by $V_{min}$, the minimum achieved by the lattice amplitude, one concludes that if  $V_{min}<V<V_{thr}$, the condition
$N_*(V)=N_A$ cannot be satisfied, what in the respective dynamical problem must result in delocalizing transition. If however $V_{min}>V_{thr}$
adiabatic increase of the lattice amplitude   will (approximately) restore the initial form of the mode.

\begin{figure}[h]
\epsfig{file=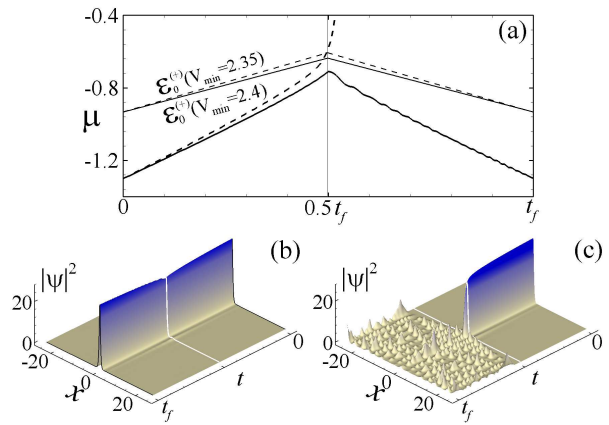,width=8cm} \caption{(a) Dynamics of the chemical potential $\mu(t)$ for $V_{min}=2.4$ (lower solid line) and $V_{min}=2.35$
(lower dashed line). The upper lines show the  respective changes of the upper edges $\cE_0^{(+)}$ of the semi-infinite gap. In (b) and (c) the
corresponding dynamics of the density profiles are shown for $V_{min}=2.4$ and $V_{min}=2.35$ respectively.  In (b) and (c) by lines we show  $t=
t_f/2$ where $V(t)=V_{min}$. Here $V_{ini}=3$, $G=0.85$, $t_f=1000$, and $N=26.25$. } \label{fig:dyn_si_bcd}
\end{figure}


The described method of finding the threshold value of the lattice amplitude was based on the static arguments. Hence the real dynamical value of
$V_{thr}$ can be slightly different, even when the change of the potential is adiabatic. Therefore we provided dynamical study of the transition
using  the following variation of the potential amplitude: $V(t)= V_{min}+(V_{ini}-V_{min}) |1-2t/t_f|$, with $t_f$ being final time of simulations
and $V(0)=V(t_f)=V_{ini}$. The minimal value $V_{min}$ is achieved at $t_f/2$. We notice, that considering a BEC in a nonlinear lattice created due
to the Feshbach resonance the  mentioned variation of the linear potential can be achieved by changing the intensity of  laser beams creating the
linear lattice subject to constant external field originating Feshbach resonance. In boson-fermion mixtures the suggested situation can also be
realized by simultaneous change of the transverse confinement of the gas, the latter affecting the effective nonlinearity together with the light
beams creating the linear periodic potential (the details as well as examples with particular boson-fermion mixtures can be found in
Ref.~\cite{SBKK}).

A typical result of the simulations is shown
in Fig.~\ref{fig:dyn_si_bcd}~(a)-(c). While in panels (b) and (c) we show the two possible scenarios: absence of delocalization (panel (b)) and
delocalizing transition (panel (c)), in the panel (a) we illustrate the evolution of the chemical potential, computed at each moment of time as $\mu=(E+E_{nl})/N$ where
\begin{eqnarray}
E=\int  \left( |\psi_x|^2+{\cal U}(x)|\psi|^2+\frac 12\int  {\cal G}(x)|\psi|^4\right)dx
\end{eqnarray}
is the energy of the condensate and $E_{nl}=\frac 12\int  {\cal G}(x)|\psi|^4dx$ is the energy of the two-body interactions. As it is clear, $\mu(t)$
has physical meaning only during the stage of the evolution until the mode is localized and follows adiabatic change of the potential. That is why
the dashed curve in Fig.~\ref{fig:dyn_si_bcd}~(a) is broken at some time corresponding to the delocalizing transition.

To describe the phenomenon in terms of the energy, we recall that for a stationary solution
\begin{eqnarray}
\label{NE}
\mu N=E+E_{nl}\quad\mbox{and}\quad
N=\frac{\partial E_{nl}}{\partial \mu}.
\end{eqnarray}
Thus, in the semi-infinite gap, a minimum of $N$ as a function of $\mu$ implies maximum of the energy of a localized mode: $E_*=E(\mu_*)=\max
E(\mu)$. Indeed, from the fact that the minimum of $N(\mu)$ is achieved at $\mu=\mu_*$ and that a given branch $E$ is uniquely determined by $N$ (and
hence by $\mu$), it readily follows that $\partial E (\mu_*)/\partial \mu=0$. Next, differentiating twice the first of equations in (\ref{NE}) with
respect to $\mu$ twice and using the second relation, one obtains
\begin{align}
\label{ENN}
    \frac{\partial^2E}{\partial \mu^2}=\frac{\partial N}{\partial \mu}+\mu\frac{\partial^2N}{\partial \mu^2}.
\end{align}
Since in the semi-infinite gap $\mu<0$ one deduces from (\ref{ENN}) that a local minimum of the number of particles with respect to the chemical potential corresponds to the local maximum of the energy of the stationary mode. We notice that there  exists a significant difference in the energetic properties of the modes in semi-infinite and in domains where the chemical  potential is positive (they correspond to finite gaps): in the last case $N_*$ corresponds to the local minimum of the energy with respect to the chemical potential.

The dynamical results depicted in Fig.~\ref{fig:dyn_si_bcd} show the predicted delocalizing transition for $V_{min}<V_{thr}$. The threshold value of the potential amplitude found from the dynamical simulations remarkably well matches
the one found from the arguments based on the static consideration. If the minimal value of the potential $V_{min} >V_{thr}$ after increase of the potential amplitude the localized modes is almost identically restored. Such behavior together with the analysis of the
behavior of the chemical potential which  approaches the gap upper edge $\cE_0^{(+)}$ as $V$ decreases [see Fig.~\ref{fig:dyn_si_bcd}~(a)], strongly support the
physical picture of the delocalizing transition suggested above.

\section{Nonlinearity induced delocalization}
\label{sec:nl_trans}

In the case of periodically varying interactions, increase of $G$ can also result in delocalizing transition, because it may change the sign of
$\chi_0^{(+)}$ (see Fig.~\ref{fig:chi}). Such transition occurs at a constant amplitude of the linear lattice. As an example, we mention that such
adiabatic change of the average nonlinearity can be achieved by changing of the boson-boson s-wave scattering length $a_s$ in the boson-fermion
mixture by external magnetic field through the Feshbach resonance~\cite{BLK} (what is possible since usually the resonant magnetic fields for
boson-boson and boson-fermion interactions are different). Moreover, an interesting way of generating delocalizing transition described in this
section comes from the fact that in the quasi-1D situation, we are considering, the quantity $N$, here referred to as a number of particles, is
related to the authentic number of atoms ${\cal N}$ by the relation $N=(4 a_s d/(\pi a^2)){\cal N}$ (here $a$ is the transverse linear oscillator
length of the bosons). This means that the effective nonlinearity can be also controlled by the lattice period. Increase of  $G$, i.e. the
delocalizing transition, can be achieved by the decrease of the lattice constant $d$, which in its turn is manipulated by the angle between the laser
beams producing the lattice~\cite{BK,OM}.

\begin{figure}[h]
\epsfig{file=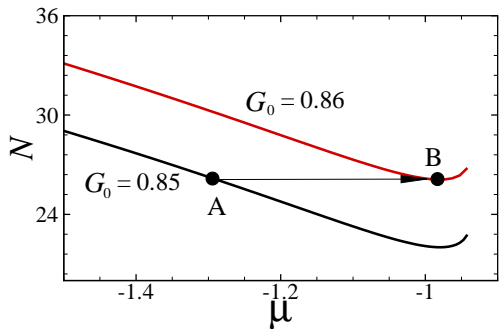,width=7cm} \caption{$N$ {\it vs}  $\mu$ for the semi-infinite gap for $G_{0}=0.85$ and $G_{0}\approx 0.86$. The arrow AB
corresponds to $N=26.25$. } \label{fig:dynG_si_a}
\end{figure}

The threshold value $G_{thr}$ at which the transition occurs can be found in a way similar to one we used to find $V_{thr}$. It is based on the
requirements (\ref{crit}) and is schematically depicted in Fig.~\ref{fig:dynG_si_a} where we show a shift of the lowest branch of the solution
subject to change of the average nonlinearity $G$. Since the number of particle $N_A$ remains constant, adiabatic increase of $G$ corresponds to the
shift along the line A$\to$B (Fig.~\ref{fig:dynG_si_a}), provided the initial modes is chosen in the point A. The critical value $G_{thr}$ is found
form the relation $N_*(G_{thr})=N_A$. Thus, designating by $G_{max}$ the maximal value of $G$, we conclude that delocalizing transition occurs if
$G_{max}>G_{thr}$ and does not occur otherwise.

\begin{figure}[h]
\epsfig{file=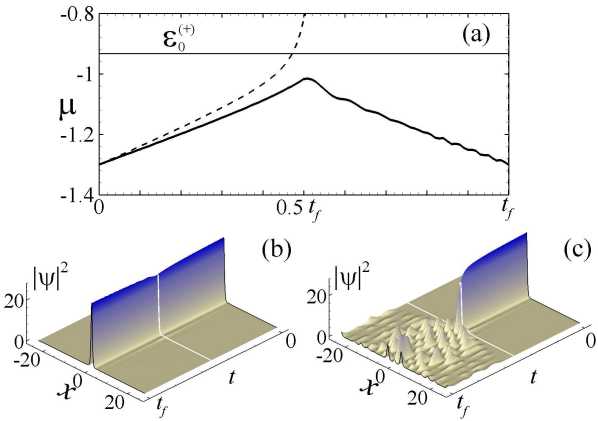,width=8cm} \caption{(a) Dynamics of $\mu$ for $G_{max}=0.86$ (lower solid line) and $G_{max}=0.862$ (dashed line). Now
$\cE_0^{(+)}=-0.937$ (upper solid line).  Dynamics of the density profiles  for $G_{max}=0.86$ (panel b) and $G_{max}=0.862$ (panel c). In (b) and
(c) by lines we show the moment $t= t_f/2$. Here $G_{ini}=0.85$, $V=3$, $t_f=1000$, and $N=26.25$.} \label{fig:dynG_si_bcd}
\end{figure}

In order to test these arguments in the dynamical scheme we use the following dependence of the nonlinearity on time: $G(t)=
G_{max}+(G_{ini}-G_{max}) |1-2t/t_f|$, i.e. $G$ initially increases until $t=t_f/2$ and then decreases to its initial value $G_{ini}$. The results of
the simulations are shown in Fig.~\ref{fig:dynG_si_bcd}.
In panel (a)
one can observe dynamics of the chemical potential which at initial stages of evolution
approaches the  gap edge and then either returns to its initial value (no delocalization, solid line) or looses its meaning at the delocalizing
transition (dashed line). Panels  (b) and (c) show explicit behavior of the spatial profiles of the modes when the  delocalizing transition is
induced by the nonlinearity with   $G_{max}>G_{thr}$, and when the transition is absent, i.e. for $G_{max}\leq G_{thr}$. As above the analysis of the
evolution of the chemical potential strongly supports the suggested physical interpretation of the phenomenon.

\section{Conclusion}

In the present paper we have shown that delocalizing transition can be observed in one-dimensional periodic systems with specific spatial dependence
of the nonlinearity, which guarantees stability of the Bloch states at the upper edge of a semi-infinite gap or at the both edges of a finite gap.
The physics of the transition is based on prohibition of existence of small-amplitude gap solitons bifurcating form the band edge. The delocalizing
transition can be originated and controlled by variation of either the lattices amplitude, or lattice period, or by change of the average nonlinearity, the latter achievable by means of magnetic of optical Feshbach resonance. In all the cases the phenomenon manifests itself in impossibility of restoring a localized shape of the mode if the change of the respective parameter goes beyond a critical value. The critical values of either the lattice amplitude or average (effective) nonlinearity can be computed with help of the suggested algorithm based on the analysis of the lowest branches of the localized modes.

The described phenomenon is not limited to the BEC applications, and should be also observable on other physical systems. In particular, we mention that in the tight-binding approximation the nonlinear Gross-Pitaevskii equation with linear and nonlinear lattices is reduced to a discrete nonlinear Schr\"{o}dinger equation with nonlocal nonlinear interactions, and thus the latter is probably the best candidate for observing one more realization of delocalizing transitions in one-dimension.

\acknowledgments

Authors are grateful to V. M. P\'erez-Garc\'ia for fruitful discussions.
YVB and VAB were supported by the FCT grants SFRH/PBD/20292/2004 and SFRH/PBD/5632/2001. VVK
acknowledges support of the Secretaria de Stado de Universidades e Investigaci\'on (Spain) under the grant
SAB2005-0195. The work was supported by the FCT and European program FEDER under the  grant POCI/FIS/56237/2004.

\end{document}